# Probing flavor changing top-charm-scalar interactions in $e^+e^-$ collisions

David Atwood,[a] Laura Reina,[b] and Amarjit Soni[b]

[a]Department of Physics, Stanford Linear Accelerator Center, Stanford, CA 94305, U.S.A.
[b]Physics Department, Brookhaven National Laboratory, Upton, NY 11973

**Abstract**: Due to the very large mass of the top-quark, probing the flavor changing top-charm-scalar vertex is clearly very important. Fortunately the largeness of $m_t$ endows a unique signature to the resulting reaction i.e., $e^+e^- \to t\bar{c}(\bar{t}c)$ that should be helpful in identification of such events. A two Higgs doublet model, without natural flavor conservation, is used to give an illustrative estimate for the rate for these reactions.

Due to its very large mass ($m_t \simeq 176$ GeV [1]) the top quark is expected to hold important clues to many outstanding issues in Particle Physics. The huge mass scale of $m_t$ suggests that we reexamine our theoretical prejudices about the existence of flavor changing scalar interactions (FCSI), especially the ones involving the top. Absence of flavor changing neutral currents (FCNC) at low energy does not necessarily forbid large FCNC at high mass scales [2]–[7]. Probing the top-charm or top-up flavor changing vertex consequently deserves a special attention. Fortunately, the heaviness of the top also facilitates experimental searches for such interactions especially in $e^+e^-$ collisions due to the clean environment that they offer. The top is so much heavier than the charm quark that when it is produced, via FCSI, in the two body reaction $e^+e^- \to t\bar{c}$ (or $\bar{t}c$), at moderate center of mass energies, the $t$ or $\bar{t}$ take up energies appreciably greater than half the total energy leading to a highly distinctive, "kinematical", signature. At $\sqrt{s}$ in the range of about 200 to 300 GeV (i.e., $m_t \lesssim \sqrt{s} \lesssim 2m_t$) detection of a $t$ or a $\bar{t}$ would clearly signify that it is produced singly (rather than in pairs). At center of mass energies above (but not far above) the pair production threshold the $t(\bar{t})$ produced via $e^+e^- \to t\bar{c}(c\bar{t})$ would have energy appreciably greater than half of the beam energy. For example, at $\sqrt{s} = 400$ GeV, $t(\bar{t})$ resulting from $e^+e^- \to t\bar{c}(c\bar{t})$ would have energy about 238 GeV rather than 200 GeV. This of course also implies that the invariant mass of the jet containing the charm quark must be close to zero. Although the excess energy of the top and the zero mass of the jet opposite to the top are necessarily related, it (i.e., the masslessness) provides an additional handle in reducing backgrounds. These kinematic features should be helpful in identifying such events. The distinctiveness of the $t\bar{c}$ signal in $e^+e^-$ is in sharp contrast to other flavor changing reactions (such as $e^+e^- \to b\bar{s}$) at high energies which are very difficult to search for experimentally.

The feasability of the experimental detection prompts one to explore the theoretical expectations for the flavor changing top-charm signals. In models with a non-minimal Higgs sector, e.g., in two Higgs doublet models, FCSI arise readily at tree level. Following Glashow and Weinberg [8] it was the practice for a long time to prevent tree level FCSI, or to implement natural flavor conservation (NFC), by imposing discrete symmetries. The rationale for this originally came from the severe suppression of FCNC processes as evidenced in e.g., $K^0 - \bar{K}^0$ oscillations, $K_L \to \mu^+\mu^-$ etc. This led to the subdivision of the commonly used two Higgs doublet models [9] with NFC into two types: 1) those in which the $u$ and $d$ type of quarks get masses from the same Higgs doublet or 2) those in which they get masses from different doublets i.e., Model I and Model II respectively. However, as was first emphasized by Cheng and Sher [2], it is rather natural to expect the Yukawa couplings for the FCSI to be related to the masses of the fermions participating at the vertices. The stringent experimental constraints against the existence of tree level FCSI involving the light quarks of the first two families can then be automatically satisfied without the need for imposing a discrete symmetry. Such a framework, which is sometimes referred to as the third type [5] of two Higgs doublet model [i.e., Model III], leads naturally to enhanced effects involving the top quark. We will use such a simple two Higgs doublet model [6] to explore some of the experimental consequences.

For self containment we will give a very brief review of the model [6] first. Since there is no global symmetry that distinguishes the two doublets, we will assume that only one of them ($\phi_1$) develops a vacuum expectation value ($v/\sqrt{2}$) and the second one ($\phi_2$) remains unbroken. The physical spectrum consists of two charged, $H^\pm$, and three neutral spin 0 bosons, $h^0$, $H^0$, which are scalars, and $A^0$ a pseudoscalar:

$$\begin{aligned}
H^0 &= \sqrt{2}[(\text{Re}\,\phi_1^0 - v)\cos\alpha + \text{Re}\,\phi_2^0 \sin\alpha] \\
h^0 &= \sqrt{2}[-(\text{Re}\,\phi_1^0 - v)\sin\alpha + \text{Re}\,\phi_2^0 \cos\alpha] \\
A^0 &= \sqrt{2}(-\text{Im}\,\phi_2^0)
\end{aligned} \quad (1)$$

The masses of the five neutral and charged spin 0 bosons, $m_H$, $m_h$, $m_A$, and $m_\pm$, as well as the mixing angle $\alpha$ are free parameters of the model. The Yukawa couplings to quarks are [6]

$$\begin{aligned}
\mathcal{L}_y^Q &= \lambda_{ij}^U \bar{Q}_i \tilde{\phi}_1 U_j + \lambda_{ij}^D \bar{Q}_i \phi_1 D_j + \xi_{ij}^U \bar{Q}_i \tilde{\phi}_2 U_j \\
&+ \xi_{ij}^D \bar{Q}_i \phi_2 D_j
\end{aligned} \quad (2)$$

As usual the first two terms here are used to give masses to the quarks and to define mass eigenstates. $\xi_{ij}^U$, $\xi_{ij}^D$ are the 3 × 3 matrices which monitor the strength of the flavor-changing neutral scalar vertices. These parameters are of course free in the model and have to be determined from experiments.

Of special importance to this work are the parameters $\xi_{tt}$ and $\xi_{tc}$. Two simple ansatz that we find interesting are the Cheng-Sher ansatz (CSA) [2, 3, 6]:

$$\xi_{ij} \sim \frac{\sqrt{m_i m_j}}{v} \qquad (3)$$

and the sum-rule ansatz (SRA):

$$\xi_{ij} \sim \frac{m_i + m_j}{2v} \qquad (4)$$

Clearly the SRA is expected to give higher rates for the FC transitions relevant to this work. Note also that some of the additional parameters (such as some of the Higgs masses) can be constrained by using these couplings in the context of low energy processes (e.g., $D^0$-$\bar{D}^0$ oscillations). However, given the degree of arbitrariness in the model we wish not to pursue that direction here. For definiteness, for now, we will content ourselves with the use of this model with only the CSA for calculating the contributions to $e^+e^- \to t\bar{c}$, $c\bar{t}$ and to other flavor changing transitions to one loop order [10].

It is useful to recall that, at one loop level, in the SM flavor changing reactions such as $e^+e^- \to t\bar{c}$, $c\bar{t}$ do arise too. The corresponding loop graphs have the generic feature that charge (-1/3) quarks appear in the loop, for charge $\frac{2}{3}$ quarks ($t$, $c$) to be produced in the final state. This feature endows the amplitude to be proportional to the square of the mass of the charge (-1/3) quark participating in the loop. In addition all the amplitudes suffer CKM suppression. Consequently the rates for $t \to c\gamma$, $cZ$ and $e^+e^- \to t\bar{c}$ are severely suppressed in the SM [11].

In a two Higgs doublet model with the usual discrete symmetry imposed to implement NFC, flavor changing transitions (e.g. $t \to c\gamma$, $cZ$ and $e^+e^- \to t\bar{c}$) again arise to one-loop order [12]. However this model retains key features of the SM (i.e., proportionality to the mass square of charge (-1/3) quarks and CKM suppression). So although the rates could be somewhat bigger than in the SM they are not large enough for experimental observation.

As a specific example $BR(t \to cZ)$ is $\simeq 10^{-13}$ in the SM [11]. In two Higgs doublet models with NFC (i.e., Model I or II) it ranges from $10^{-14}$ to $10^{-9}$ [12]. In Model III (i.e., without NFC), relevant to this paper, $BR(t \to cZ) \sim 10^{-9}$–$10^{-7}$ [6, 10]. Thus in models without flavor conservation, the branching ratios and cross sections for FC transitions can be larger by about two orders of magnitude compared to models with NFC.

Now, the amplitude for $e^+e^- \to t\bar{c}(\bar{t}c)$, via $Z$ or $\gamma$ exchange, is given by:

$$M^{Z,\gamma} = \frac{1}{16\pi^2} \bar{e}\,\gamma_\alpha(a_e^{\gamma,Z} + b_e^{\gamma,Z}\gamma_5)\,e\,\Pi^{\gamma,Z} \qquad (5)$$

$$\bar{t}\left[\gamma^\alpha(A^{\gamma,Z} + B^{\gamma,Z}\gamma_5) + i\frac{q_\mu}{m_t}\sigma^{\alpha\mu}(C^{\gamma,Z} + D^{\gamma,Z}\gamma_5)\right]c$$

where $a_e^{\gamma,Z}$ and $b_e^{\gamma,Z}$ are SM couplings of the $\gamma$ and $Z$ to the electron, $q = p_t + p_c$ and $\Pi^{\gamma,Z}$ is the $\gamma, Z$-propagator respectively. $A^{\gamma,Z}$ etc. are the form factors at the $tc$ vertex, being calculated to the one loop order [13]. Assuming CSA we will parametrize the Yukawa couplings as:

$$\xi_{ij} = g\frac{\sqrt{m_i m_j}}{m_W}\lambda \qquad (6)$$

and for now we will ignore CP-violation and take $\lambda$ to be real. From eqns.(5) and (6) we see that the $(t, c)$ form factors ($A^{\gamma,Z}$, etc.) scale as $\lambda^2$. Consequently, the cross section for $e^+e^- \to t\bar{c}$ would go as $\lambda^4$ and would be a very sensitive probe of the important parameter $\lambda$ that characterizes the overall strength of FCSI.

We present the numerical results in Figs.(1)–(3) for the total cross section normalized to the $\mu^+\mu^-$ cross section via one-photon exchange, i.e.

$$R^{tc} \equiv \frac{\sigma(e^+e^- \to t\bar{c} + \bar{t}c)}{\sigma(e^+e^- \to \gamma \to \mu^+\mu^-)} \qquad (7)$$

We will split the discussion into four cases:

1. all scalars are roughly degenerate i.e., $m_h \simeq m_A \simeq m_\pm \simeq M_s$ where $M_s$ is the common scalar mass.

2. $m_h$ is "light" (denoted by $M_\ell$) and $m_A \simeq m_\pm = 1$ TeV.

3. $m_A = M_\ell$ and $m_h \simeq m_\pm = 1$ TeV.

4. $m_\pm = M_\ell$ and $m_h \simeq m_A = 1$ TeV.

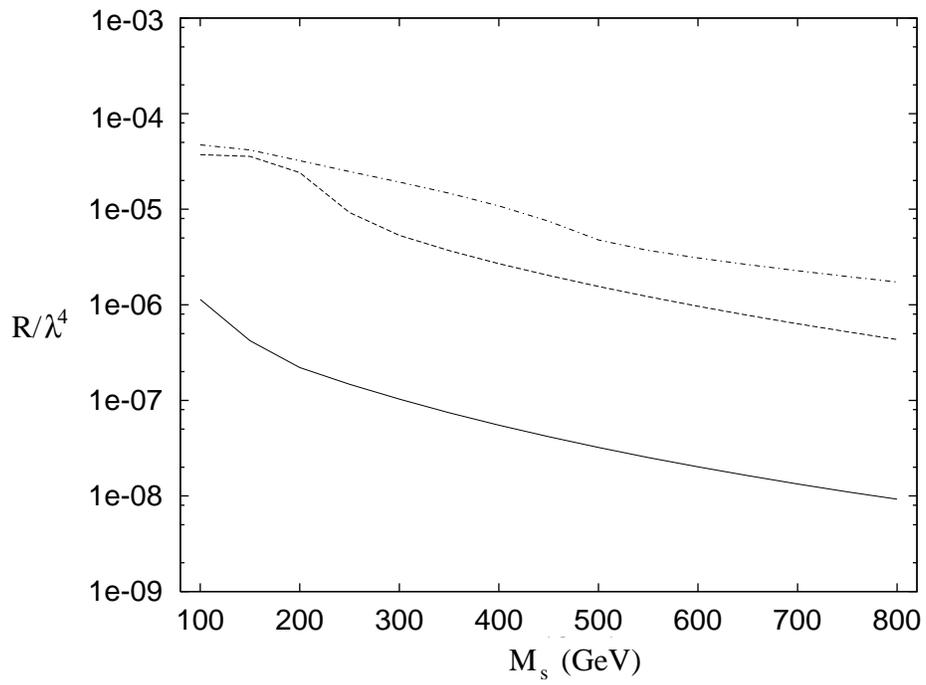

Figure 1: $R^{tc}/\lambda^4$ vs. the common scalar mass, $M_s$ for $\sqrt{s} = 200$ (solid), 500 (dashed), and 1000 GeV (dot-dashed).

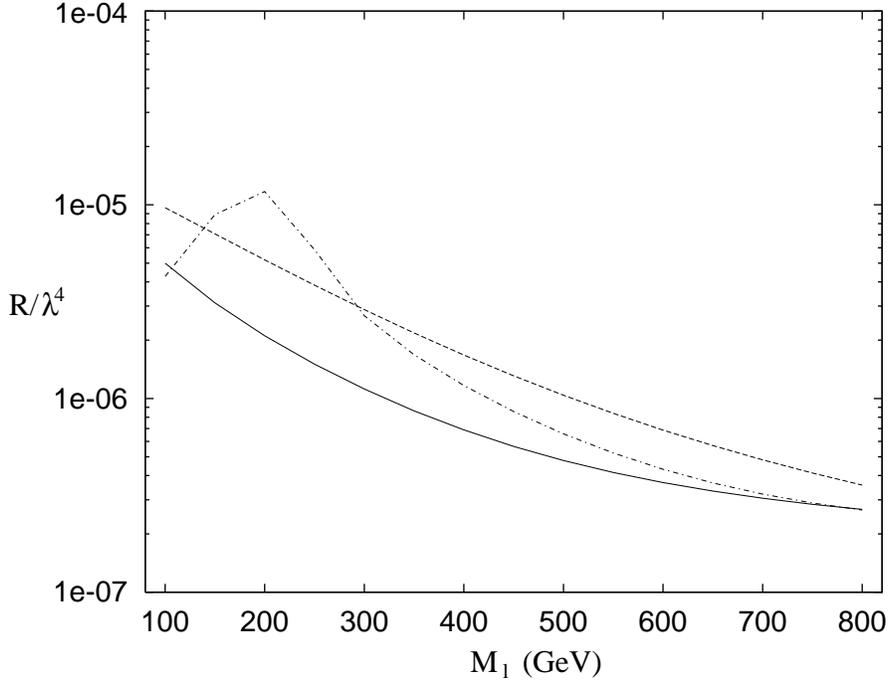

Figure 2: $R^{tc}/\lambda^4$ vs. the light scalar mass, $M_\ell$, for $\sqrt{s} = 500$ GeV. Case 1 (solid) i.e., $m_h = M_\ell$ and $m_A \simeq m_\pm = 1$ TeV; case 2 (dashed) i.e., $m_A = M_\ell$ and $m_h \simeq m_\pm = 1$ TeV; case 3 (dot-dashed) i.e., $m_\pm = M_\ell$ and $m_h \simeq m_A = 1$ TeV.

Fig.(1) shows $R^{tc}/\lambda^4$ as a function of $M_s$ for the case when all scalar masses are roughly degenerate (i.e., case 1). The rates are displayed for three values of the beam energy, i.e., 200, 500 and 1000 GeV. Note that for $M_s \sim 200$ GeV and $\lambda \sim 1$, $R^{tc}$ can be a few times $10^{-5}$ which should be at the detectable level since it is reasonable to expect $10^6$–$10^7$ $\mu^+\mu^-$ events in a year of running. Fig.(2) shows the three possibilities (cases 2, 3 and 4) depending on which of the scalars is the lightest. In this figure we take $\sqrt{s} = 500$ GeV. The peak in $R^{tc}/\lambda^4$ for case 4, when charged Higgs is lightest, occurs due to the appearance of the threshold for pair production of $H^\pm$. Fig.(3) shows $R^{tc}/\lambda^4$ as a function of $\sqrt{s}$ again for cases 2–4. From these figures, we see that for each of the four cases enumerated above $R^{tc}/\lambda^4$ can be about $10^{-5}$. Thus we can expect experiments to be able to constrain $\lambda \lesssim 1$, for scalar

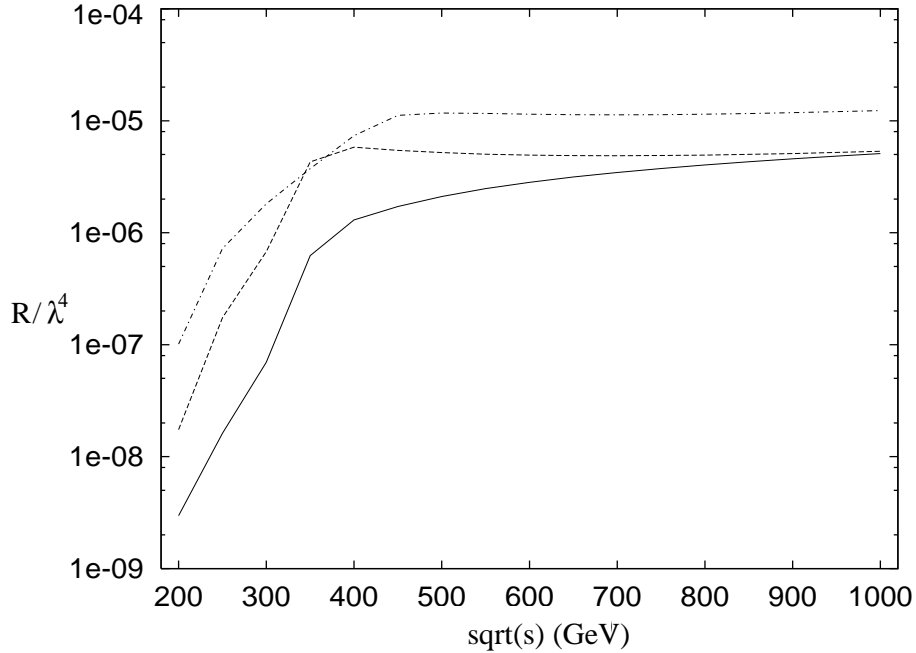

Figure 3: $R^{tc}/\lambda^4$ vs. $\sqrt{s}$ with $M_\ell = 200$ GeV for case 1 (solid), case 2 (dashed) and case 3 (dot-dashed). (See also caption for Fig.(2))

masses of a few hundred GeVs.

To summarize, in this paper, we have emphasized the importance of searching for flavor-changing-scalar interactions via the reactions $e^+e^- \to t\bar{c}$. We stressed that the experimental signal is very clean and that mild extensions of the Standard Model, say with an extra Higgs doublet [6], complemented with the popular Cheng-Sher [2] ansatz can lead to measurable effects. There is no experimental basis for assuming the absence of tree level flavor-changing neutral currents at the mass scale of $m_t$ [2]-[7]. Consequently their vigorous search is strongly advocated.

One of us (A.S.) is greatful to George Hou and Tony Sanda for discussions. This work was supported by U.S. Department of Energy contracts DE-AC03-76SF00515 (SLAC) and DE-AC-76CH0016 (BNL).

[13] Note that the expressions for most of these form factors are directly related to the ones given in Ref.[6] (for $\alpha = 0$ the corresponding Feynman diagrams are those of Fig.(1) in that reference). However, in our case, $A^\gamma$, $B^\gamma \neq 0$ as we are dealing with an off-shell photon. We have used the opportunity to recalculate all the form factors. In Ref.[10] we will give explicit expressions for them and also their numerical values and we will compare them with Ref.[6].